# Optical Probing of Electronic Interaction between Graphene and Hexagonal Boron Nitride

Gwanghyun Ahn,[1] Hye Ri Kim,[2] Taeg Yeoung Ko,[1] Kyoungjun Choi,[2] Kenji Watanabe,[3] Takashi Taniguchi,[3] Byung Hee Hong[2,4] and Sunmin Ryu[1]*

[1]Department of Applied Chemistry, Kyung Hee University, Yongin 446-701, Korea

[2]SKKU Advanced Institute of Nanotechnology (SAINT), Center for Human Interface Nano Technology (HINT) and Department of Chemistry, Sungkyunkwan University, Suwon 440-746, Korea

[3]National Institute for Materials Science, 1-1 Namiki, Tsukuba, 305-0044, Japan

[4]Department of Chemistry, Seoul National University, Seoul 151-747, Korea

*E-mail: sunryu@khu.ac.kr

**Abstract**

**Even weak van der Waals (vdW) adhesion between two-dimensional solids may perturb their various materials properties owing to their low dimensionality. Although the electronic structure of graphene has been predicted to be modified by the vdW interaction with other materials, its optical characterization has not been successful. In this report, we demonstrate that Raman spectroscopy can be utilized to detect a few % decrease in the Fermi velocity ($v_F$) of graphene caused by the vdW interaction with underlying hexagonal boron nitride (hBN). Our study also establishes Raman spectroscopic analysis which enables separation of the effects by the vdW interaction from those by mechanical strain or extra charge carriers. The analysis reveals that spectral features of graphene on hBN are mainly affected by change in $v_F$ and mechanical strain, but not by charge doping unlike graphene supported on $SiO_2$ substrates. Graphene on hBN was also found to be less susceptible to thermally induced hole doping.**

**Keywords: graphene, boron nitride, Raman spectroscopy, 2D band, electronic coupling**

Numerous studies since the first graphene field effect transistors[1] have revealed the adverse effects of the popular $SiO_2$ substrates on the device performance and materials properties. In general, graphene supported on silica suffers mobility decrease due to substrate-induced ripples,[2-4] scattering from charge impurities[5-8] and surface optical phonons[8,9] of the substrates. The rough surface morphology of commercially available silica substrates leads to structural deformation of the supported graphene generating nanometer-scale ripples[2,10,11] and charge puddles.[7,12] Deformed graphene is also more vulnerable to chemical attacks[13-15] and develops strong p-type charge doping caused by ambient oxygen molecules.[11]

Hexagonal boron nitride (hBN), a chemically inert and thermally robust[16] dielectric material with optical bandgap of 5.97 eV,[17] was the first alternative substrate to remedy the silica-induced effects providing improved carrier mobility and decreased native charge density due to its crystalline nature and lack of surface dangling bonds.[18] When supported on hBN, graphene is flatter with an order-of-magnitude smaller roughness[18] and slight lattice mismatch of 1.7%,[19] suggesting better structural quality than that on silica substrates. Moreover, its high optical phonon frequency with dielectric properties comparable to those of silica makes hBN suitable for high temperature or electric-field applications.[20] Heterostructures like graphene/hBN formed by stacking 2-dimensional materials not only improve device performance but also allow new phenomena and functionalities to be discovered. Tunneling through artificial graphene bilayers sandwiching a nm-thick hBN layer obeys exponential dependence on the thickness of the spacer[21] and the resulting field-effect tunneling transistor showed an improved on/off switching ratio of ~50.[22] By controlling charge density in one graphene layer of the sandwich, Anderson localization was observed in the other graphene layer leading to metal-insulator transition.[23] Moreover, the van der Waals (vdW) interaction, despite being weak, has been predicted to lift degeneracy of the neighboring two C atoms and open up a bandgap in a Bernal-stacked graphene/hBN heterostructure,[19] whereas no gap was realized in experiments[18,24] due to random stacking.[25] The vdW interlayer interaction is also manifested in stacking-dependent moire patterns in graphene/hBN,[26] and modulation in electronic band structures in Bernal- and random-stacked graphene bilayers.[27]

Raman spectroscopy has been widely used in graphene study to characterize charge density,[28,29] mechanical strain,[30-37] mix[38] of both, and temperature[39] as well as number of layers,[40] stacking[41] and defects.[40,42] Its Raman 2D band has served as a spectroscopic fingerprint in distinguishing single-layer (1L) graphene from Bernal-stacked multilayers.[40] Random-stacking in twist graphene bilayers also modifies the electronic structure near K points in the Brillouin zone inducing "twist angle"-dependent reduction in Fermi velocity ($v_F$),[43] non-dispersive D band,[44] and G band enhancement.[45, 46] Unlike graphene-graphene homo-stacks, however, optical characterization of electronic coupling in hetero-stacks made of graphene and other materials has been rare despite the rising interest.[47] Because of the high sensitivity of the 2D peak frequency ($\omega_{2D}$) to $v_F$, in particular, even a slight change in the electronic structure of graphene through the vdW interaction will influence the Raman spectral features, which should enable quantification of the electronic perturbation.[43] Understanding and separating this coupling between electronic and nuclear degrees of freedom are also important in establishing "graphene metrology" by Raman spectroscopy[38, 48] where users have to rely on the two Raman peaks to quantify the aforementioned multiple factors.

Herein we demonstrate that the interlayer interaction modifies the linear dispersion of graphene on hBN, but not on silica, leading to ~7 cm$^{-1}$ upshift in $\omega_{2D}$, which translates into ~3% decrease in $v_F$. Unlike on silica, the native charge density of graphene is very low and annealing-induced hole doping is greatly reduced on hBN. This study shows that even weak interlayer interactions can influence Raman spectra of graphene in contact with other materials, and thus complements the Raman spectroscopic graphene metrology mainly reserved for strain[31, 32, 48] and charge doping.[28, 29, 38]

**Results and Discussion**

Hetero-stacks of graphene and hBN were prepared by a simple mechanical transfer (see Methods for details). First, thin hBN flakes were deposited on Si substrates with 285 nm-thick SiO$_2$ layer through mechanical exfoliation[1] of hBN crystals.[17] Graphene grown on Cu foils by chemical vapor deposition method (CVD) was deposited onto the SiO$_2$/Si substrates with hBN flakes using the standard etching and transfer methods.[49] Figure 1a presents the optical micrograph of sample **G1**, which consists of a thin hBN layer (~20x8 μm$^2$) and SiO$_2$ area both covered with graphene, denoted 1L$_{BN}$ and 1L$_{SiO2}$, respectively. Since multilayer domains (>1 μm$^2$) can be easily noticed in the optical micrograph (area marked by yellow arrows in Fig. 1a & 1b), optical microscopy was used to select samples with high coverage of graphene and small (<1%) areal fraction of multilayer graphene which complicates interpretation of Raman spectra.[43] The AFM image in Fig. 1b, obtained within the yellow box in Fig. 1a, indeed revealed that graphene covers most (> 95%) of the scanned area with the rest corresponding to cracks or holes in the graphene sheet. Areas covered with multilayer graphene are scarcely found only near the torn holes marked by the yellow arrow, indicating that the CVD growth is limited to single layer.[49] Figure 1b and 1c also show that the transfer step generated wrinkles or folds in graphene. The detailed AFM image in Fig. 1c, however, confirms that the transferred graphene is quite flat except the wrinkles suggesting good contact with the substrates. Since the graphene area corresponding to the wrinkles turned out to be less than 1% of the whole from the surface area analysis of Fig. 1c, their contribution to the Raman spectra should also be negligible (see Supporting Information, Fig. S2). The thickness of the hBN layer is 3.4 ± 0.2 nm as shown in Fig. 1d presenting the line profile averaged in the yellow rectangle in Fig. 1c. Height histograms in Fig. 1e confirm that the surface of bare hBN is much flatter than that of $SiO_2$ substrates:[18] the standard deviation for 7 nm-thick hBN is 90 pm mostly due to instrumental noise,[18] whereas that for $SiO_2$ is 280 pm.

Figure 2a presents two Raman spectra each obtained respectively from $1L_{SiO2}$ and $1L_{BN}$ of Fig. 1a. The spectrum from $1L_{SiO2}$ shows the two prominent Raman peaks, G and 2D respectively at ~1590 and ~2689 cm$^{-1}$, indicating substantial p-type charge doping as will be discussed below. The disorder-related D band also appears at ~1350 cm$^{-1}$ and the D-to-G peak height ratio ($I_D/I_G$) was found to be ~0.10 throughout the sample. Since $I_D/I_G$ of graphene transferred onto bare $SiO_2$/Si substrates was ~0.05, we attribute the additional D intensity to the wrinkles, cracks and holes aggravated during the transfer of graphene by the presence of hBN flakes and adhesive residues on hBN/$SiO_2$/Si. To further confirm the thickness of the CVD-grown graphene, we quantified the amount of C atoms using the G peak area ($A_G$) of mechanically exfoliated graphene which follows a quasi-linear relation between its $A_G$ and thickness[50] (Supporting Information, Fig. S1). $A_G$ of the CVD-grown 1L graphene turned out to lie within 10% from that of exfoliated 1L graphene, corroborating the thickness assignment. However, $A_G$ of CVD-grown random-stacked 2L graphene in Fig. S1 was equal to or significantly larger than that of exfoliated 2L graphene. The enhancement in $A_G$, due to the singularities in the joint density of states,[45] limits reliable thickness characterization in random-stacked multilayers. It is to be noted that the intensity, lineshape and linewidth ($\Gamma_{2D}$) of 2D also vary nonlinearly as a function of the twist angle in random-stacked bilayers[45, 46] and that $\Gamma_{2D}$ and $A_{2D}/A_G$ are much less useful in determining thickness than $A_G$ (Fig. S1).

The spectrum of $1L_{BN}$ in Fig. 2a shows another sharp peak at 1366 cm$^{-1}$, originating from the $E_{2g}$ phonon mode of hBN crystal.[51] The spectral details of the hBN peak were obtained by separation from the D peak through a curve fitting as shown in Fig. 2a. The Raman map for the hBN peak area ($A_{BN}$) in Fig. 2b matches well with the optical micrograph and AFM images in Fig. 1. Whereas the G and 2D peaks are also most prominent in $1L_{BN}$, their spectral details are distinct from those of $1L_{SiO2}$. The G peak frequency ($\omega_G$ ~ 1584cm$^{-1}$), ~6 cm$^{-1}$ lower than that from $1L_{SiO2}$, is more closer to the intrinsic value of graphene ($\omega_G^o \sim 1581.5$ cm$^{-1}$),[38] which can also be seen in the $\omega_G$-map in Fig. 2c. Additionally, the spectra reveal the linewidth of the G peak ($\Gamma_G$) and 2D-to-G peak area ratio ($A_{2D}/A_G$) are larger for 1L$_{BN}$ than 1L$_{SiO2}$. These spectral differences, occurring throughout the sample as shown in the Raman maps of Fig. 2e & 2f, can be explained by reduced charge doping[18] in 1L$_{BN}$ as will be discussed below and are consistent with the scanning tunneling microscopy study of CVD graphene on hBN.[26] However, we note that the change in $\omega_{2D}$ from its intrinsic value ($\omega_{2D}^o \sim 2677$ cm$^{-1}$)[38] is unusually high ($\Delta\omega_{2D} \sim 11$ cm$^{-1}$) and cannot be solely attributed to mechanical strain or charge doping, since $\Delta\omega_G$ is only $\sim 2.5$ cm$^{-1}$ and thus $\Delta\omega_{2D}/\Delta\omega_G$ is larger than 4.[38]

To interpret the anomalous behavior of $\omega_{2D}$, we employed the analysis recently proposed by J. Lee *et al.*,[38] which distinguishes the effects of the two most influential factors in Raman spectra of graphene, mechanical strain[30-37] and charge doping.[28, 29] The Raman peak frequencies ($\omega_G$, $\omega_{2D}$) of graphene under tensile (compressive) stress will move from the intrinsic value of strain-free and charge-neutral graphene, **O**($\omega_G^o$, $\omega_{2D}^o$),[38] along the **e$_T$** (**e$_C$**) vector representing tensile (compressive) strain as shown in the inset of Fig. 3a. Hole doping will move ($\omega_G$, $\omega_{2D}$) along the **e$_H$** vector in the inset as the data from an electrical gating measurement[52] show in Fig. 3a (red solid line). Using strain ($\varepsilon$) and charge density ($n$) values marked on the **e$_T$** and **e$_H$** axes, any given ($\omega_G$, $\omega_{2D}$) can then be vector-decomposed into $\varepsilon$ and $n$.[38] For instance, two groups of ($\omega_G$, $\omega_{2D}$) points[38] obtained from a graphene sample mechanically exfoliated from graphite (Fig. 3a) clearly reveal its pristine state with varying native strain ($-0.1\% < \varepsilon < 0.4\%$) but negligible charge density (brown squares) and hole-doped state ($n \sim 1.4 \times 10^{13}$ cm$^{-2}$) induced by thermal annealing (brown triangles).

When projected onto the ($\omega_G$, $\omega_{2D}$) space in Fig. 3a, the Raman data of two samples **G1** and **G2** processed in the same conditions with similar hBN thickness are grouped into two distinct regions, each for 1L$_{SiO2}$ (circles) and 1L$_{BN}$ (crosses), respectively. All the samples studied showed the same grouping behavior (see Supporting Information). As previously mentioned regarding Fig. 2, Fig. 3 clearly shows that 1L$_{SiO2}$ areas suffer hole doping of varying density ($n < 4 \times 10^{12}$ cm$^{-2}$) with **G2** less doped than **G1**. Figure 3 further reveals that the spread in ($\omega_G$, $\omega_{2D}$) due to strain in 1L$_{SiO2}$ areas is much smaller than that due to varying charge density. Now we note that 1L$_{BN}$ shows a very different spectral behavior. The data points for 1L$_{BN}$ are centered around (1583.3, 2687.9) cm$^{-1}$ for **G1** and (1583.9, 2688.5) cm$^{-1}$ for **G2** in the forbidden zone[38] which cannot be reached by a linear combination of strain (**e$_T$** or **e$_C$**) and hole doping (**e$_H$**). We attribute this anomaly in 1L$_{BN}$ to modification of graphene's electronic structure caused by vdW interaction with hBN. More specifically, modulation in the dispersion of $\pi$ or $\pi^*$ bands, approximated as change in v$_F$,[27] leads to change in observed $\omega_{2D}$, since D phonon mode of different wave vector will be selected by the double resonance processes.[43] Since $\omega_G$ originating from the $E_{2g}$ zone center phonon should not be affected to a first-order approximation, the electronic modulation causing reduction in $v_F$ should move ($\omega_G$, $\omega_{2D}$) along $\mathbf{e_{FVR}}$ (denoting Fermi velocity reduction) as shown in the inset of Fig. 3a.

However, a given ($\omega_G$, $\omega_{2D}$) cannot be decomposed along the three unit vectors unambiguously, since all three vectors in 2-dimension cannot be independent of each other. Thus separation of the contributions from the three factors requires knowledge of at least one of the three. In Fig. 3b, we present $A_{2D}/A_G$ which decreases rapidly as increasing $|n|$.[53] It can be seen that the ratios for $1L_{BN}$ (5.6 ± 0.2 for **G1**; 6.0 ± 0.2 for **G2**) are large and close to those for charge-neutral graphene denoted by the green dot (6.2 ± 0.2) in Fig. 3b while that for $1L_{SiO2}$ is significantly smaller and widely spread just like $\omega_G$ in Fig. 3a. Since $A_{2D}/A_G$ is very sensitive to low level of charge density,[53] we conclude that $n$ of $1L_{BN}$ areas is very small and insignificant compared to $n \sim 2\times10^{12}$ cm$^{-2}$ for **G2**'s $1L_{SiO2}$. Assuming that the spectral changes for $1L_{BN}$ occurred only along -$\mathbf{e_T}$ ($\mathbf{e_C}$) or $\mathbf{e_{FVR}}$, the change in $\omega_{2D}$ along $\mathbf{e_{FVR}}$ ($\Delta\omega_{2D}^{FVR}$) can be estimated to be 7.2 and 6.5 cm$^{-1}$ for **G1** and **G2**, respectively. The analysis also leads to the fact that both $1L_{BN}$ areas are slightly compressed with $\varepsilon \sim -0.1\%$. The estimated degree of strain, however, is subject to whether the strain is uniaxial or biaxial.[38] Whereas graphene grown on Cu foils through CVD is likely to be under biaxial stress due to isotropic differential thermal expansion of Cu,[54, 55] it was shown that the substrate-induced strain (or charge doping) is largely removed when transferred onto other substrates.[56] In addition, graphene may undergo further mechanical deformation during wet etching and transfer processes using polymer supports.[49] Although the nature of the native strain in **G1** and **G2** samples cannot be further revealed, it is to be noted that graphene mechanically exfoliated onto silica substrates is mostly under randomly oriented uniaxial stress,[38] implying that random mechanical perturbation like mechanical exfoliation or physical transfer favors uniaxial stress unlike the isotropic thermal perturbation.

In Fig. 4a, we investigated $1L_{BN}$ regarding spectral changes due to thermal stress which causes $O_2$-induced hole doping and compression in mechanically exfoliated graphene on SiO$_2$ substrates as shown in Fig. 3a by brown symbols.[11, 38] Upon thermal annealing at 400 °C for 2 hours, $1L_{SiO2}$ of sample **G3** showed a drastic change in ($\omega_G$, $\omega_{2D}$), which corresponds to $\Delta n \sim 1\times10^{13}$ cm$^{-2}$ confirming emergence of the strong hole doping.[38] In contrast, the spectral change of $1L_{BN}$ was much less and associated $\Delta n$ is only $\sim 3\times10^{12}$ cm$^{-2}$. $A_{2D}/A_G$ ratios in Fig. 4b also confirm that $1L_{BN}$ is much less susceptible to the thermal perturbation. The distribution of peak frequencies and area ratios increased by annealing can be attributed to the spectral inhomogeneity caused by structural deformation or *in-situ* reactions at elevated temperature.[38]

Our study shows that hBN induces much less charge doping in graphene upon thermal annealing than SiO$_2$ substrates. Whereas exact mechanistic understanding has yet to be made, the thermally induced hole doping in graphene on SiO$_2$ is caused by ambient oxygen molecules in the presence of water molecules.[11, 13] The molecular doping is also apparently connected to thermal generation[11] of microscopic ripples caused by conformal adhesion[57] to rough substrates or slipping-rippling[38] due to negative thermal expansion of graphene. Since hBN is highly flat and also has negative thermal expansion coefficient[58] unlike SiO$_2$, thermal rippling is expected to be much less on hBN. Moreover, hydrophobic hBN surface should contain or attract less water which enhances the O$_2$-induced hole doping than hydrophilic SiO$_2$ abundant with surface silanol groups.[59] Since the charge doping is activated by thermal treatment at as low as 100°C,[38] alternative substrates like hBN will be useful in future graphene applications which require reliable control of charge density or electrical conductivity.

The current study also reveals noticeable effects of hBN on the Raman spectra of graphene. Because of the random relative orientations and translations,[27] however, the interlayer interaction in our 1L$_{BN}$ samples is expected to be smaller than that for graphene in good stacking registry with hBN like AA' and AB, for which theory predicted adhesion energy of 20 ~ 30 meV/C atom.[19] It is also to be noted that the adhesion energy is significantly lower than the interlayer cohesive energy in graphite (61 meV/C atom)[60] or adhesion energy between graphene and SiO$_2$ substrates (74 meV/C atom).[61] Despite the weak vdW interaction, however, the observed $\Delta\omega_{2D}^{FVR}$ for 1L$_{BN}$ is significant enough to estimate the degree of modification in the electronic structure. Theory predicted that interlayer coupling in twist bilayer graphene preserves the linear dispersion near K points but with reduced $v_F$ which is dependent on the twist angle.[27] Using Raman spectroscopy, Ni *et al.* determined $\Delta v_F/v_F$, reduction in $v_F$ of twist bilayer graphene, which varied from -2 to -6% for several samples with unknown twist angles.[43] Similarly, one can estimate the change in 1L$_{BN}$ using $\Delta v_F/v_F = 0.00449\ \Delta\omega_{2D}^{FVR}$ which has been modified from what Ni *et al.* derived considering different $\omega_{2D}^o$ and excitation photon energy: The values of $\Delta\omega_{2D}^{FVR}$ for 1L$_{BN}$ lead to $\Delta v_F/v_F$ of -3.2 and -2.9% for **G1** and **G2**, respectively.

Now we discuss the effects of vdW interaction with SiO$_2$ substrates on the two Raman modes. Despite many Raman spectroscopy studies on graphene supported on SiO$_2$/Si substrates, the effects of vdW interaction on $v_F$ and phonon frequency have not been clearly understood because of the overwhelmingly large spectral variations caused by native charges and strain.[38, 62] Recently, however, J. Lee *et al.* showed that ($\omega_G$, $\omega_{2D}$) of charge-neutral graphene supported on SiO$_2$ nicely follows the **e$_T$** line, which indicates that the spectral variation is exclusively due to native strain.[38] Despite the significant interfacial adhesion,[61] their data show no apparent movement along **e$_{FVR}$** within their experimental uncertainty of 1

cm$^{-1}$, implying negligible change in v$_F$ and thus $\omega_{2D}$. This may be attributed to the fact that SiO$_2$ is in amorphous phase thus not providing periodic perturbation to the band structure. Furthermore, the partial suspension on SiO$_2$ substrates[38, 63, 64] may reduce the effects of the underlying substrates. On the other hand, the vdW interaction may change the force constants of the Raman modes directly. Direct observation of the change, however, is not straightforward due to the large native spectral variations in graphene supported on SiO$_2$ substrates. Viewing the fact that $\omega_G$ of freestanding graphene [38, 65] is almost identical to that of Bernal-stacked graphite which is essentially vdW-type complex of graphene, one may predict that $\omega_G$ is not strongly affected by vdW interaction with SiO$_2$ substrates which have similar interaction energy as the interlayer cohesion energy in graphite. J. Lee *et al.*'s data also suggest that $\omega_G$ is not directly affected by the vdW interaction with SiO$_2$ substrates.[38] We also note the Raman spectroscopy work[66] by C. Lee *et al.* on single layer of semiconducting MoS$_2$, where the frequencies of E$^1_{2g}$ and A$_{1g}$ Raman modes were found to be highly homogeneous unlike graphene. Exploiting freestanding MoS$_2$, they showed that the frequencies of the two Raman modes are not affected (within 0.3 cm$^{-1}$) by the presence of SiO$_2$ substrates.

Although there have been many Raman spectroscopy studies on graphene with mechanical strain and extra charge carriers both mediated by underlying substrates and environment, systematic and quantitative analysis has not been performed to separate the effects of both until J. Lee *et al.*'s report.[38] For example, random stiffening of G and 2D modes observed in pristine graphene on amorphous[62] or crystalline insulators[67] was attributed to spontaneous p or n-type doping without considering native strain. The spectral changes in graphene that underwent thermal treatments were controversially interpreted as either mechanical compression[68, 69] or chemical charge doping.[11, 13, 70, 71] Some chemical treatments were considered to result in charge doping exclusively.[14, 72] Epitaxial graphene grown on 6H-SiC[56, 73-75] and Ru(0001)[76] has been claimed dominated by strain with minor charge doping. All of these systems are potentially susceptible to multiple perturbations simultaneously. In this regard, our work should provide a further refined approach in graphene metrology using Raman spectroscopy complementing the recent work[38] by J. Lee *et al*. In particular, graphene on crystalline substrates, like graphene on hBN, may be also affected by the interfacial vdW interaction in addition to strain or charge transfer, which demands careful interpretation as proposed in the current study. Despite its utility, however, our approach cannot avoid the inherent limitation that mix of more than two factors cannot be disentangled in $\omega_G$-$\omega_{2D}$ space without additional information. Furthermore, the effect of n-type charge doping on $\Delta\omega_{2D}/\Delta\omega_G$ is highly nonlinear unlike that of p-type,[38] which would complicate its separation.

**Conclusion**

In summary, we have demonstrated that weak vdW interaction between graphene and crystalline substrates can be detected by Raman spectroscopy. Whereas $\omega_G$ is not affected, $\omega_{2D}$ increases due to the decrease in the Fermi velocity of graphene caused by the adhesion on hBN. This observation establishes a simple optical method to separate the effects of the vdW interaction entangled with those of mechanical strain or charge doping. The current study also reveals that Raman spectra of graphene on hBN are mostly affected by the vdW interaction and mechanical strain, but negligibly by charge doping, which contrasts with graphene supported on $SiO_2$ substrates. The proposed analysis should serve as a fast and reliable optical probe of strain or excess charges in graphene suffering vdW interaction with underlying crystalline substrates.

**Methods**

**Preparation of graphene/hBN samples.** Using mechanical exfoliation[1] of hBN crystals,[17] thin hBN flakes were first deposited on Si substrates which were covered with 285 nm-thick $SiO_2$ layer. Then, graphene grown on Cu foils by the CVD method was deposited onto the $SiO_2$/Si substrates decorated with the thin hBN flakes using the standard etching and transfer methods.[49] The thickness of hBN flakes and morphology of the hetero-stacks were revealed by atomic force microscopy (AFM; XE-70, Park Systems). To avoid complication due to possible mechanical strain in graphene enveloping hBN flakes, flakes thinner than 7 nm were chosen for this study. Thermal annealing was carried out for 2 hours at specified temperature in a vacuum tube furnace maintained below 3 mTorr.

**Raman spectroscopy.** The Raman spectra were obtained by a home-built micro-Raman setup also detailed previously.[38] Briefly, excitation laser beam with a power of 1.5 mW operated at 514.5 nm was focused onto a spot of 0.5 μm in diameter using a 40 times objective lens with a numerical aperture of 0.6, which then collected backscattered Raman signal. Spectral accuracy was better than 1.0 cm$^{-1}$ as described in a recent report.[38] To obtain statistically meaningful data, Raman mapping was carried out in a region of >20x20 μm$^2$ per each sample by raster-scanning every 1 μm along x and y axes, thus providing more than 400 independent probe spots.

**Acknowledgments**

This work was supported by the National Research Foundation of Korea (No. 2012-053500, 2012-043136, 2012-0003059, 2011-0021972).

**Supporting information available**

Optical determination of thickness using $A_G$, surface area of wrinkles, Raman spectral analysis of additional graphene-BN samples. This material is available free of charge *via* the Internet at http://pubs.acs.org.

**Figure legends**

Figure 1. Morphology of graphene-hBN heterostack. (a) Optical micrograph of hBN/SiO$_2$/Si covered with CVD-grown graphene (sample **G1**), where 1L$_{BN}$ and 1L$_{SiO2}$ designate graphene areas contacting hBN and SiO$_2$, respectively. (b) Non-contact AFM height image (9x9 μm$^2$) obtained from the area within the yellow square in (a). (c) Non-contact AFM height image (2x2 μm$^2$) obtained from the area within the white square in (b). (d) Height profile averaged from the yellow rectangle in (c). The thickness of the hBN flake, defined by the height difference between the two shaded regions in (d), is 3.4 ± 0.2 nm. (e) Height histograms of bare hBN (red circles) and SiO$_2$ substrates (blue circles). Roughness defined by standard deviation for Gaussian distribution in solid curves was 90 and 280 pm, respectively. The blue square in (a) marks the area where the Raman maps shown in Fig. 2 were obtained. The yellow arrows in (a) & (b) indicate areas where graphene was ruptured and folded during the transfer process.

Figure 2. Raman spectra and maps of graphene-hBN heterostack. (a) Raman spectra of 1L$_{BN}$ and 1L$_{SiO2}$ (**G1**). D, G and 2D denote Raman peaks, respectively, originating from D mode, G mode and overtone of D mode. The peak denoted BN is due to E$_{2g}$ phonon mode of hBN crystal. The detailed spectra (black squares for 1L$_{SiO2}$ and red circles for 1L$_{BN}$) separately shown for the D peak region reveal the presence of BN peak along with D peak for 1L$_{BN}$, with both peaks well described by double Lorentzian functions (orange and green lines). (b) Raman map for BN peak area (A$_{BN}$). (c) Raman map for G peak frequency

($\omega_G$). (d) Raman map for 2D peak frequency ($\omega_{2D}$). (e) Raman map for G peak linewidth ($\Gamma_G$). (f) Raman map for 2D-to-G peak area ratio ($A_{2D}/A_G$). Mapping was carried out by raster scanning the blue squared region (20x20 μm$^2$) in Fig. 1(a) with each pixel corresponding to an area of 1x1 μm$^2$. The dotted black lines in (b) ~ (f) represent the boundary of the hBN flake shown in Fig. 1(a).

Figure 3. Raman spectral analysis of graphene-BN heterostack. (a) Correlation between $\omega_G$ and $\omega_{2D}$ of **G1** (red symbols) and **G2** (blue symbols). Crosses and open circles represent 1L$_{BN}$ and 1L$_{SiO2}$, respectively. Brown squares and triangles, obtained respectively from pristine and thermally annealed graphene/SiO$_2$ (Ref. 38), are shown for comparison. Inset: The arrows labeled **$e_T$**, **$e_C$**, **$e_H$** and **$e_{FVR}$** represent the trajectories of **O**($\omega_G$, $\omega_{2D}$) affected respectively by tensile strain, compressive strain, hole doping and vdW interlayer interaction leading to Fermi velocity reduction. The tick labels for $\varepsilon$ on the **$e_T$** axis in (a) are given assuming uniaxial strain (Ref. 36) and those for *n* and $\Delta v_F/v_F$ along **$e_H$** and **$e_{FVR}$** are based on Ref. 52 and Ref. 43, respectively. (b) $A_{2D}/A_G$ of **G1** and **G2** as a function of $\omega_G$. The green dot and solid line represent average $A_{2D}/A_G$ of freestanding graphene (Ref. 38) with uncertainty marked by the error bars and dotted lines. The black circles and error bars represent respectively average and standard deviation values for 1L$_{SiO2}$ data, whereas orange squares and error bars correspond to those for 1L$_{BN}$ data.

Figure 4. The effects of thermal annealing on strain and charge doping. (a) Correlation between $\omega_G$ and $\omega_{2D}$ of **G3** obtained before (blue symbols) and after (red symbols) thermal annealing for 2 hours at 400 °C in vacuum. Crosses and open circles represent 1L$_{BN}$ and 1L$_{SiO2}$, respectively. (b) $A_{2D}/A_G$ of **G3** as a function of $\omega_G$. The green dot and solid line represent average $A_{2D}/A_G$ of freestanding graphene (Ref. 38) with uncertainty marked by the error bars and dotted lines. The black circles and error bars represent respectively average and standard deviation values for 1L$_{SiO2}$ data, whereas orange squares and error bars correspond to those for 1L$_{BN}$ data.

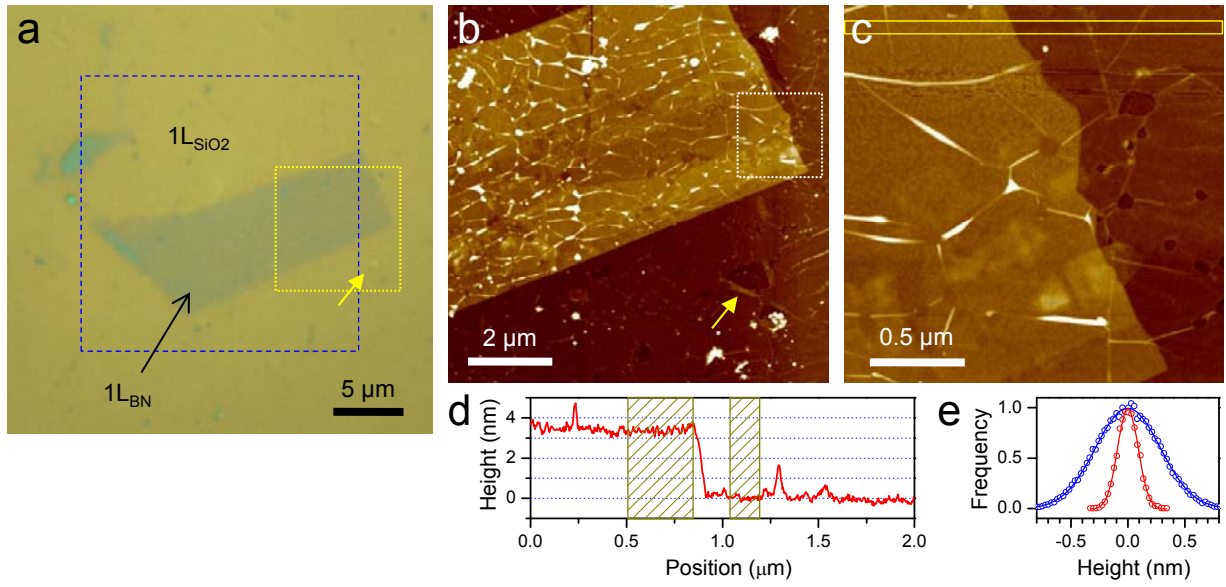

Figure 1. G. Ahn et al.

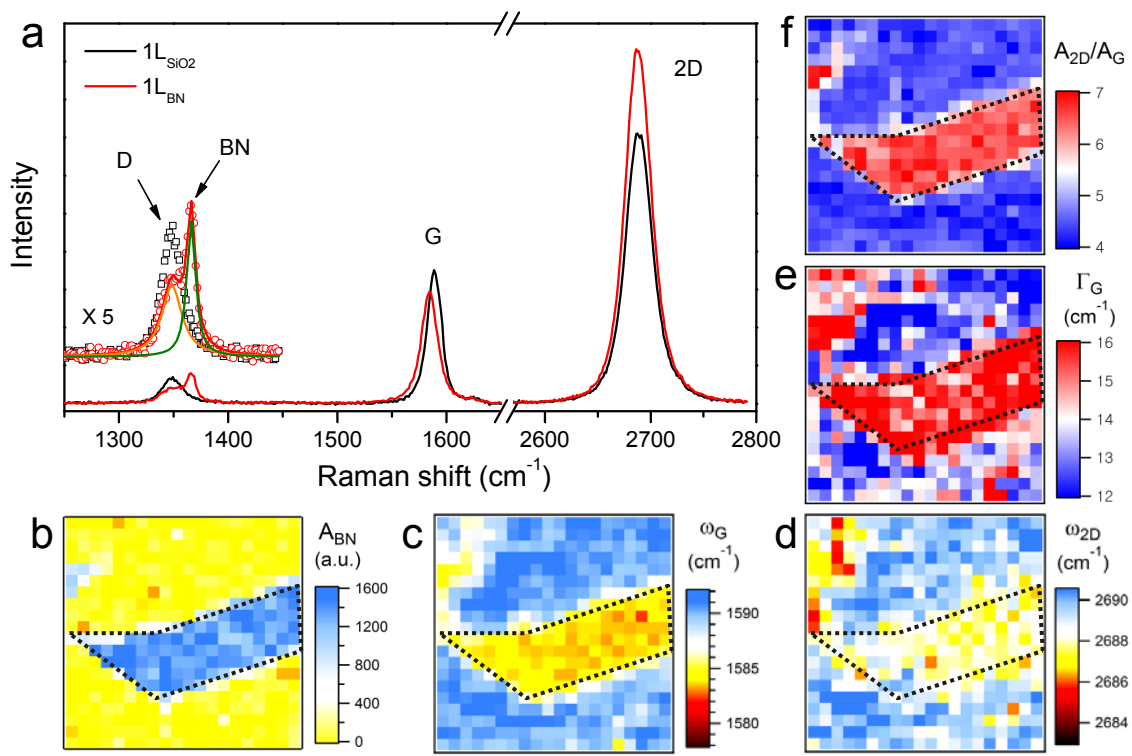

Figure 2. G. Ahn et al.

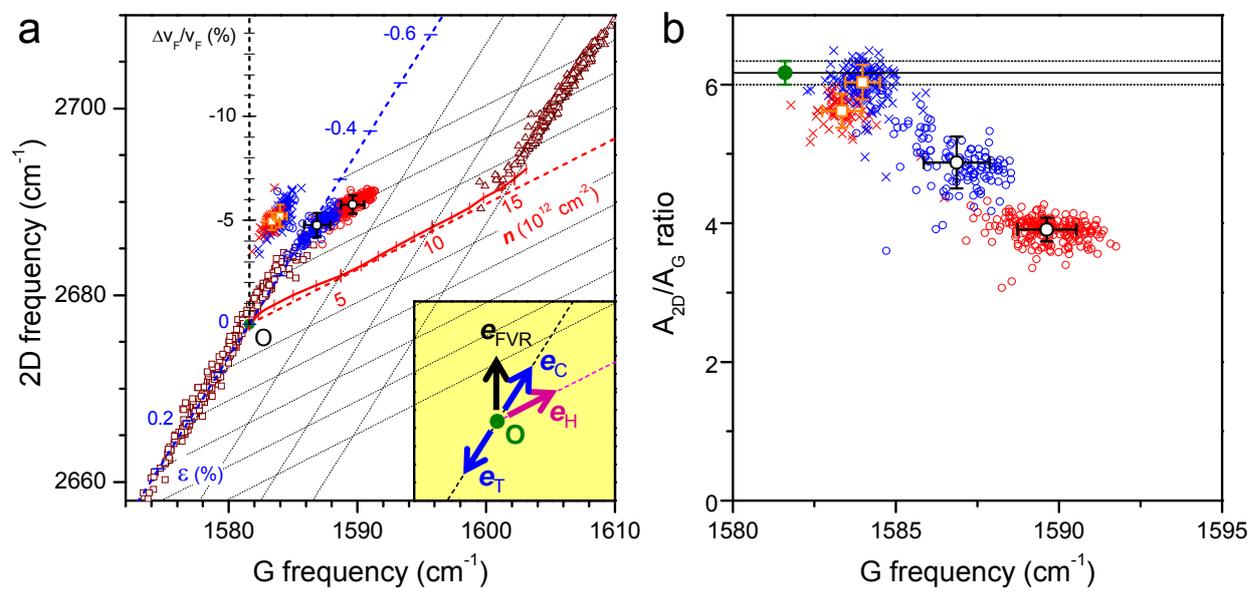

Figure 3. G. Ahn et al.

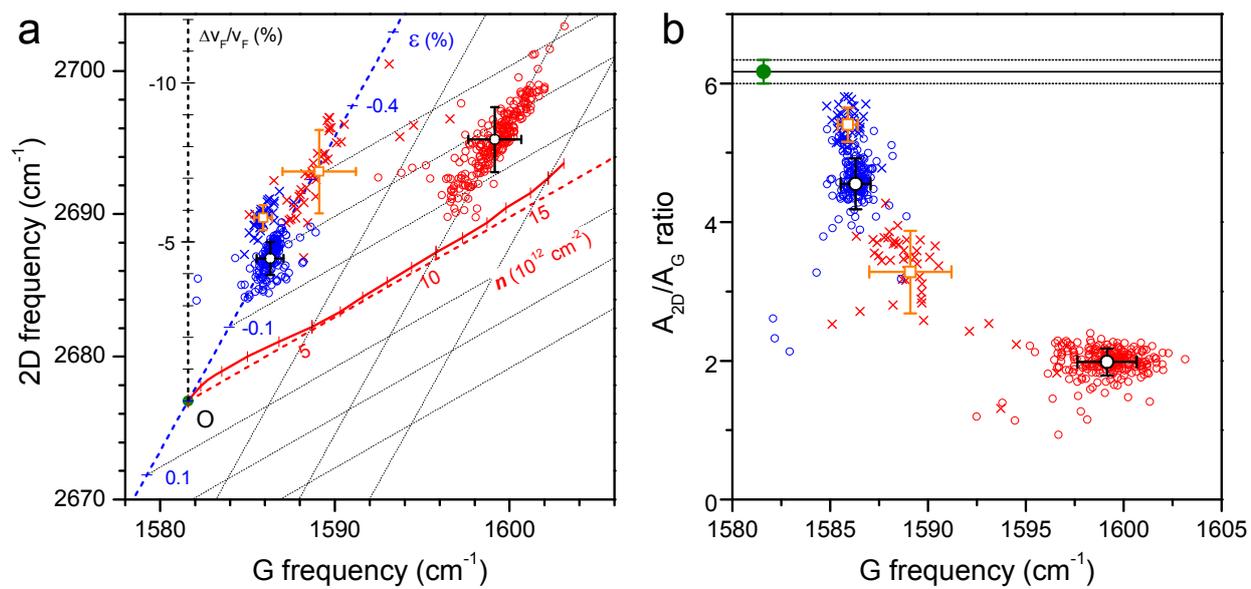

Figure 4. G. Ahn et al.